\documentclass[superscriptaddress,onecolumn]{revtex4-2}
\usepackage{graphicx}
\usepackage{newtxtext}
\usepackage{newtxmath}

\newcommand{\RomanNumeralCaps}[1]

\usepackage[english]{babel}
\usepackage{amssymb}
\usepackage{amsmath}
\usepackage{stmaryrd}
\usepackage{siunitx}
\usepackage{array}
\usepackage{paralist}
\usepackage{enumitem}
\usepackage[version=4]{mhchem}
\usepackage{xcolor}
\usepackage{float}
\usepackage{csquotes}
\usepackage[colorlinks=true,urlcolor=blue,citecolor=blue]{hyperref}

\newcommand{\bff}{{\boldsymbol f}}
\newcommand{\md}{{\mathrm{d}}}
\newcommand{\pll}{{\parallel}}
\newcommand{\bu}{{\boldsymbol u}}

\newcommand{\bp}{{\boldsymbol p}}
\newcommand{\bx}{{\boldsymbol x}}

\newcommand{\re}{\textrm{Re}}
\newcommand{\uvc}[1]{\hat{\bf{#1}}} 

\begin{document}
\title{Dynamics of flexible filaments in oscillatory shear flows}

\author{Francesco Bonacci}
\thanks{F.B. and B.C. contributed equally to this work.}
\affiliation{PMMH, CNRS, ESPCI Paris, Université PSL, Sorbonne Université, Université de Paris, F-75005, Paris, France}
\author{Brato Chakrabarti}
\thanks{F.B. and B.C. contributed equally to this work.}
\affiliation{Center for Computational Biology, Flatiron Institute, New York, New York 10010, USA}
\author{David Saintillan}
\affiliation{Department of Mechanical and Aerospace Engineering, University of California San Diego, La Jolla, California 92093, USA}
\author{Olivia du Roure}
\affiliation{PMMH, CNRS, ESPCI Paris, Université PSL, Sorbonne Université, Université de Paris, F-75005, Paris, France}
\author{Anke Lindner}
\email[Correspondence:]{ anke.lindner@espci.psl.eu}
\affiliation{PMMH, CNRS, ESPCI Paris, Université PSL, Sorbonne Université, Université de Paris, F-75005, Paris, France}

\begin{abstract}
The fluid-structure interactions between flexible fibers and viscous flows play an essential role in various biological phenomena, medical problems, and industrial processes. Of particular interest is the case of particles freely transported in time-dependent flows. This work elucidates the dynamics and morphologies of actin filaments under oscillatory shear flows by combining microfluidic experiments, numerical simulations, and theoretical modeling. Our work reveals that, in contrast to steady shear flows, in which small orientational fluctuations from a flow-aligned state initiate tumbling and deformations, the periodic flow reversal allows the filament to explore many different configurations at the beginning of each cycle. Investigation of filament motion during half time periods of oscillation highlights the critical role of the initial filament orientation on the emergent dynamics. This strong coupling between orientation and deformation results in new deformation regimes and novel higher-order buckling modes absent in steady shear flows. The primary outcome of our analysis is the possibility of suppression of buckling instabilities for certain combinations of the oscillation frequency and initial filament orientation, even in very strong flows. We explain this unusual behavior through a weakly nonlinear Landau theory of buckling, in which we treat the filaments as inextensible Brownian Euler-Bernoulli rods whose hydrodynamics are described by local slender-body theory.
\end{abstract}
\maketitle

\section{Introduction\label{sec:intro}}

\noindent The dynamics of slender fibers in viscous flow is key to our understanding of many complex phenomena encountered in diverse areas ranging from biology to physics and engineering~\citep{DuRoure2019}. Advances in microfluidics have made it possible to probe various fluid-structure interaction problems. Recent progress has centered around the dynamics of freely transported filaments in steady flows. Examples include morphological dynamics in steady shear~\citep{Schroeder2005,Harasim2013,kuei2015dynamics,Liu2018,zuk2021universal} and hyperbolic extensional flows in the vicinity of stagnation points~\citep{Schroeder2003,Kantsler2012,Manikantan2015,Chakrabarti2020}. However, many biophysical processes and industrial lab-on-a-chip applications, such as cardiovascular transport~\citep{Tarbell2014}, filtration of cells and clog mitigation~\citep{Lee2018,Cheng2016}, or sorting and mixing of particles~\citep{Dincau2020} involve unsteady or time-periodic flows. Elucidating the physics behind the transport and deformation of flexible fibers in such conditions is thus of paramount importance yet remains nascent.

In the present work, we focus on studying the dynamics and morphologies of elastic Brownian filaments in oscillatory shear flow. We use individual F-actin filaments as an experimental model system. Actin is a semi-flexible biopolymer usually found in the cell cytoskeleton and plays a variety of important roles in sub-cellular processes~\citep{Pollard2003}. In physiological conditions, actin polymerizes into filaments whose length $L$ is of the order of their persistence length $\ell_p$. As a result, they possess a degree of flexibility intermediate between the limits of entropy-dominated long-chain polymers such as DNA ($\ell_p\ll L$) and of stiff, rigid rods ($\ell_p\gg L$) such as microtubules. For filaments with $L \sim \ell_p$, the bending and thermal energies have similar magnitudes, and they together determine the emergent dynamics under viscous loading. The combination of filament elasticity coupled with their rotation~\citep{Jeffery1922} in flow is central to a plethora of rich dynamics. In particular, actin filaments can undergo buckling instabilities when viscous forces overcome their bending rigidity, much like in the Euler buckling of a macroscopic elastic beam mechanically loaded at its ends. This results in a series of morphological transitions in steady shear flow \citep{Harasim2013,Liu2018,slowicka2020flexible,zuk2021universal}, is responsible for the so-called stretch-coil transition \citep{young2007stretch,Wandersman2010,Kantsler2012}, and can lead to the formation of helicoidal structures \citep{Chakrabarti2020} in compressional flow. In parallel, numerical explorations suggest that these structural instabilities strongly impact the statistical properties of tumbling dynamics under shear~\citep{Munk2006,Lang2014}, leading to deviations in classical scaling laws of flexible and stiff polymers~\citep{Schroeder2003}. These microscopic instabilities also play a role in the macroscopic rheological properties of polymeric suspensions where they give rise to shear-thinning and positive normal stress differences in shear flow \citep{Becker2001,Tornberg2004,Chakrabarti2021a}.

In such steady flows, the dynamics of a filament is primarily characterized by the dimensionless elastoviscous number $\bar{\mu}_\text{m}$ that compares the time scale of bending relaxation to the shear rate and provides an effective measure of the strength of viscous forces compared to elastic resistance. In simple steady shear flow, compressive viscous forces can lead to an Euler-like buckling instability at a critical value of $\bar{\mu}_\text{m}^c \approx 306.8$ \citep{Becker2001}. This gives rise to deformed $C$ shaped configurations. In even stronger flows, higher-order buckling modes are triggered along with a series of conformational transitions that have previously been characterized \citep{Harasim2013,Liu2018}. 

Time-periodic oscillatory flows introduce another time scale in the problem, namely the period $T$ of oscillation, making way for a variety of unexplored morphological dynamics. Here we systematically characterize filament dynamics in an oscillatory shear flow through a combination of fluorescence microscopy experiments in microfluidic channels, theoretical modeling, and Brownian dynamics simulations. Our experiments use vertical Hele-Shaw cells to drive a horizontal oscillatory flow generated by a pair of pressure controllers connected to the channel inlets. For all experimental conditions, we checked that the Womersley number $W\!o$~\citep{Womersley1995} was $W\!o\ll 1$, meaning that oscillation-induced inertial effects were negligible compared to viscous forces. As a consequence, the flow is essentially proportional to the instantaneous pressure gradient and well approximated at the length scale of a filament as a simple shear flow in the horizontal plane with $\dot{\gamma}(t) = \dot{\gamma}_m \sin{(2\pi t/T)}$. The additional time scale is described through a dimensionless number $\rho = \dot{\gamma}_m T$ that compares the maximum shear rate $\dot{\gamma}_m$ to the time period $T$ of the imposed flow.

In this work, we focus on three central questions. First, we characterize filament morphologies and possible dynamics over a half time period $T/2$ of the imposed shear flow. Our findings reveal that, in contrast to steady shear, the emergent dynamics are sensitive to initial filament orientations and are strongly affected by the time period $T$ of the flow. Second, we show that the coupling of orientational dynamics of the filaments with the flow results in novel higher-order buckling modes absent in steady shear. Finally, we find that unique to the periodic forcing of the flow is a possibility of suppression of buckling instabilities, even at maximum shear rates that exceed the buckling threshold for steady flow ($\bar{\mu}_\text{m} \gg \bar{\mu}_\text{m}^c$). We corroborate this remarkable dynamical behavior through a weakly nonlinear Landau theory of buckling. 
The paper is organized as follows: in \S~\ref{sec:problem}, we discuss the experimental set-up, the modeling framework and governing equations of the problem. We present our key findings from experiments, simulations and theory in \S~\ref{sec:results}, and we conclude in \S~\ref{sec:concl}.

\section{Experimental methods, modeling and governing equations}\label{sec:problem}

\subsection{Materials and methods} \label{sec:m&m}

\begin{figure}
  \centering
  \includegraphics[width=0.8\linewidth]{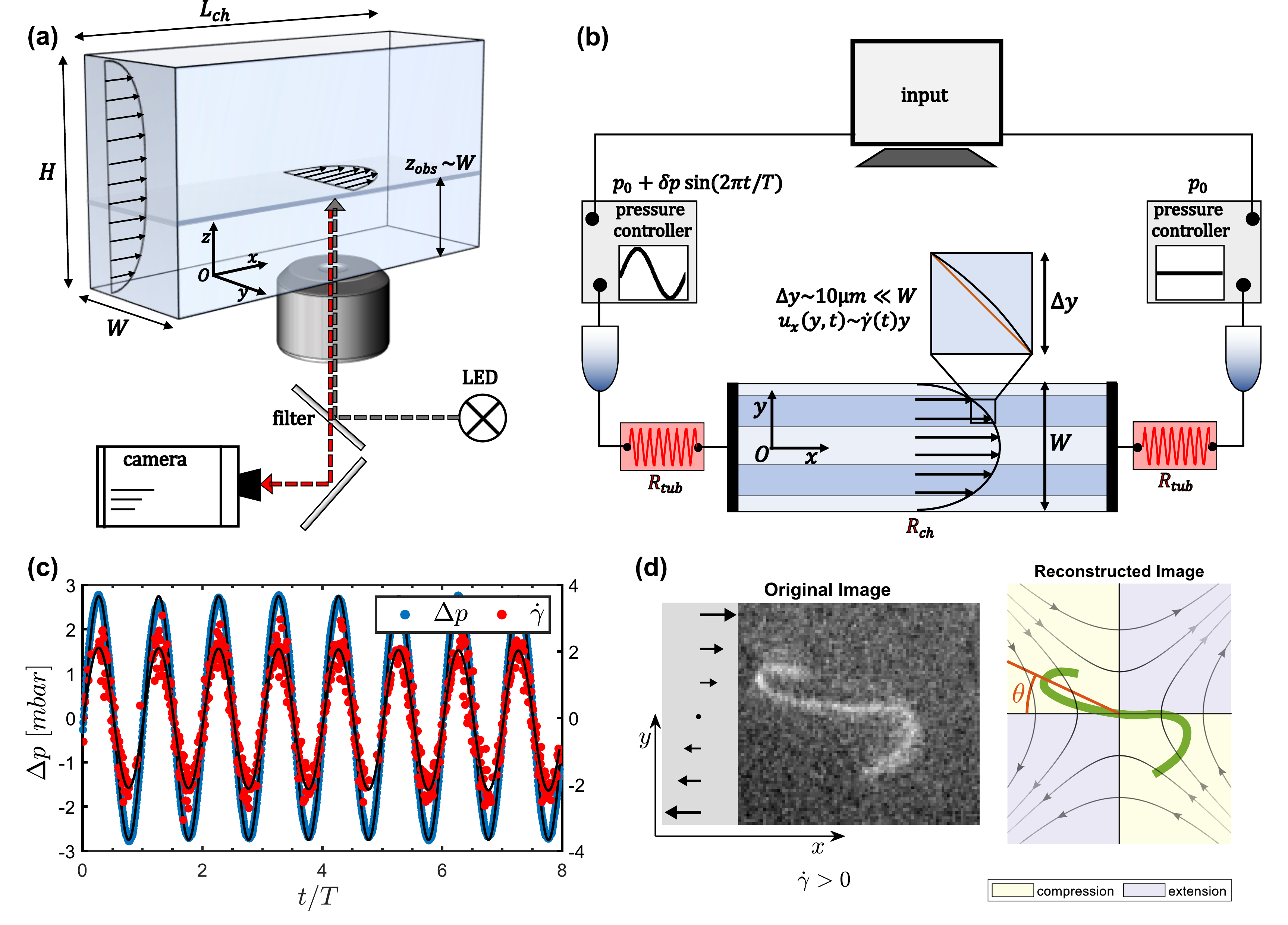}
  \caption{Sketch of the set-up used to carry out oscillatory shear flow experiments on actin filaments. {\bf{(a)}} Lateral 3D-view of the micro-channel and imaging system. {\bf{(b)}} Top view of the micro-channel along with the system for generating pressure-driven oscillatory flows. {\bf{(c)}} Imposed pressure (blue symbols) and local shear rate experienced by the filament (red symbols) in a typical oscillatory experiment. Solid black lines are sinusoidal fits. {\bf{(d)}} Raw and processed image of a filament. Overlaid to the reconstructed centerline are the compression and extension quadrants of the flow. According to our notation, the mean orientational angle is measured clockwise from the negative $x$-axis. This results in compressive viscous stresses for $0^{\circ}< \theta <90^{\circ}$, and extensile stresses for $90^{\circ}<\theta<180^{\circ}$ when the shear rate is positive, as in the case shown. }
  \label{fig:setup}
\end{figure}

We follow a well controlled and reproducible protocol for the synthesis of actin filaments, the details of which are extensively described in~\cite{Liu2018}. The filaments in our experiments are fluorescently labelled and stabilized with phalloidin.  They have typical lengths of $L \sim 5-25 \ \mu \mathrm{m}$, diameter $a \sim 8 \ \mathrm{nm}$ and persistence length $\ell_p=$\SI{17}{\micro \meter}, as measured by analysis of thermal-fluctuation-induced conformational changes~\citep{Liu2018b,Brangwynne2007}.

A sketch of the experimental set-up is shown in Fig.~\ref{fig:setup}. The experiments are conducted in an inverted microscope equipped with a LED light source (Ziess Colibri 7) for dye excitation and a high numerical aperture (NA) water-immersion objective (Zeiss 63x C-Apochromat /1.2 NA), which provides large working distances of up to \SI{280}{\micro \meter}.

The flow is driven in vertical Hele–Shaw poly(dimethylsiloxane) (PDMS) channels of length $L_{ch} =$ \SI{30}{mm}, height $H =$ \SI{500}{\micro\meter} and width $W =$ \SI{150}{\micro\meter} [Fig.~\ref{fig:setup}{\bf{(a)}}]. Thanks to the high NA of the objective, we visualize the filaments at distance $z_{obs}\simeq W = $ \SI{150}{\micro\meter} from the coverslip, where the flow profile is Poiseuille-like in the horizontal $x\hat{O}y$ plane and has a velocity plateau along the channel height~\citep{Liu2018}; this guarantees that the filaments are mainly deformed by a nearly two-dimensional shear flow in the observation plane. In order to avoid wall-interaction effects and the change of sign of $\dot\gamma$ near the channel center, we only consider filaments that are located inside the dark blue regions in Fig.~\ref{fig:setup}{\bf{(b)}}. Besides, the channel width $W$ being much larger than the typical dimension of the deformed filament ($\sim$\SI{10}{\micro \meter}), the flow gradient can be approximated as constant over that length scale, with the $y$-position determining the actual shear rate experienced by the filament~\citep{Liu2018}.

Since we observe filaments far away from the coverslip, we face two major problems. First, the intensity of the fluorescent light emitted by the filament is low. Second, a slight refractive index mismatch between the solution and the PDMS can result in optical aberrations. To circumvent these problems, we add 45.5$\%$ (wt/vol) sucrose to the actin solution, which allows us to match the refractive index of the PDMS channel ($n=1.41$). This  increases the suspension viscosity to \SI{5.6}{mPas}~\citep{Liu2018}. 

Images are acquired using a CMOS camera (HAMAMATSU ORCA flash 4.0LT) at relatively large exposure times in the range of $t_{\text{exp}} = 40-60$ \SI{}{ms}. This, together with the high NA of the objective, allows the collection of as much light as possible from the filaments, therefore enhancing image quality, but limits the maximum frequency of acquisition to $16-25$ fps {and sets an upper boundary to the maximum flow speed where filaments can be observed without image blur.} 
Images are processed using a custom-made MATLAB routine, which involves a Hessian-based multiscale filtering, noise reduction using Gaussian blurring, binarization, skeletonization and B-spline reconstruction; see Fig.~\ref{fig:setup}{\bf{(d)}} for an example.

The periodic pressure-driven flow is imposed through a pair of commercially available pressure controllers (Fuigent Lineup Flow EZ\textsuperscript{TM}) connected to the channel inlets [Fig.~\ref{fig:setup}{\bf{(b)}}]. Each unit has a typical response time of \SI{50}{ms}, a pressure drop in the range of $1-25$ \SI{}{mbar} with \SI{0.1}{mbar} precision, and is controlled by a dedicated software. In order to apply a sinusoidal flow with zero offset, one channel inlet is provided with a constant pressure of magnitude $p_0$, while a time-dependent pressure $p(t) = p_0 + \delta p\sin{\left(2\pi t/T\right)}$ is imposed in the opposite inlet ($\delta p=1.3-12$ \SI{}{mbar}). This results in a total pressure drop $\Delta p(t) = \delta p\sin{\left(2\pi t/T\right)}$.  Typical data are reported in Fig.~\ref{fig:setup}{\bf{(c)}} in blue.
 
We apply maximum flow rates in the range of $Q = 0.8 - 11$ \SI{}{nL/s}, corresponding to typical maximum velocities of $u_x = 15 - 175$ \SI{}{\micro m/s} in the observation plane, filament Reynolds numbers around $10^{-5}$ to $10^{-4}$, and maximum shear rates of $\dot\gamma_m=0.4-3.7 \ \mathrm{s}^{-1}$. In order to obtain these shear rate values, we have to reduce the pressure drop in the channel. This is achieved  as shown in Fig.~\ref{fig:setup}{\bf{(b)}} by inserting before each channel inlet a very thin tubing (length $l\approx 10$ \SI{}{cm}, inner diameter $d= \SI{75}{\micro m}$) with flow resistance $R_{tub}$ much larger than the channel resistance $R_{ch}\sim1.5 \cdot10^{12}$ \SI{}{Pa\; s/m^3}, which is estimated as $R_{ch} = 12\mu L_{ch}/[W^3H (1-0.63W/H)]$~\citep{Bruus2008}. This typically yields a total flow resistance $R_{tot} = 2R_{tub}+R_{ch}$ with $R_{tot}/R_{ch}\sim100$.  

  
Unlike in low-\re\;steady flows where only viscous effects dominate, ensuring laminar conditions, inertia can be important in a time-dependent flow and possibly lead to deviations from the steady velocity profile as well as phase shifts between the flow and the pressure gradient, even for $\re\ll 1$. We compute the Womersley number $W\!o$~\citep{Womersley1995}, a dimensionless parameter that compares the relative importance of transient inertial effects to viscous forces:
\begin{equation}
W\!o = D\left(\frac{2\pi\varrho}{T\mu}\right)^{1/2},     
\end{equation}
where $D$ is a characteristic length scale of the flow (here, taken as the largest transverse size of the channel $D\sim H$) and $\varrho$ the fluid density. In our experiments the Womersley number is set by the oscillation frequency, and decreases from $W\!o\sim0.5$ to $\sim0.15$ as we vary the period $T$ in the range of $1 - 10$ \SI{}{s}. Since in all cases $W\!o<1$, inertia is not relevant whereas viscous resistance dominates. As shown in Appendix \ref{app:A}, the use of small Womersley numbers guarantees that a steady Poiseuille-like velocity profile has time to develop during each cycle, and that the flow is nearly in phase with the instantaneous external pressure~\citep{Dincau2020}.

In this frequency regime, therefore, the time-dependent shear rate experienced by the filament essentially follows the imposed pressure. This is clearly demonstrated in Fig.~\ref{fig:setup}{\bf{(c)}}, where the reported shear rate (red symbols), reconstructed from the filament center-of-mass velocity along $x$ (refer to Appendix \ref{app:A}), is well fitted by a sinusoidal function $\dot{\gamma}(t) = \dot{\gamma}_m \sin{(2\pi t/T)}$.

\subsection{Slender-body-theory for a Brownian filament}

We choose to model the slender filaments as 1D spacecurves described by their centerline, which is identified by a Lagrangian marker $\bx(s,t)$ parametrized by arclength $s \in [-L/2,L/2]$. Hydrodynamics is captured using local slender-body theory (SBT) \citep{Tornberg2004}, in which the centerline position evolves as:
\begin{equation}
8 \pi \mu\left[\partial_{t} \bx(s, t)-\bu_{\infty}(t)\right]=-\boldsymbol{\Lambda} \cdot \bff(s, t). \label{eq:sbtheory}
\end{equation}
Here, $\mu$ is the fluid viscosity and $\bu_\infty(t) = \left(\dot{\gamma}(t) y,0,0\right)$ is the background shear flow with an oscillatory shear rate $\dot{\gamma}(t) = \dot{\gamma}_m \sin(2 \pi t/T)$. The force per unit length exerted by the filament on the fluid is modeled as $\boldsymbol{f}=B \boldsymbol{x}_{s s s s}-\left(T \boldsymbol{x}_{s}\right)_{s}+\boldsymbol{f}^{b}$, where $B$ is the bending rigidity, $T(s)$ is a Lagrange multiplier that enforces inextensibility of the filament and can be interpreted as a line tension, and $\bff^b$ is the Brownian force density obeying the fluctuation-dissipation theorem. The local mobility operator $\boldsymbol{\Lambda}$ accounts for drag anisotropy and is given by
\begin{equation}
\boldsymbol{\Lambda} \cdot \boldsymbol{f}=\left[(2-c) \mathbf{I}-(c+2) \boldsymbol{x}_{s} \boldsymbol{x}_{s}\right] \cdot \boldsymbol{f},  \label{eq:mobility}
\end{equation}
where $c = \ln(\epsilon^2 e) < 0$ is an asymptotic geometric parameter depending on the aspect ratio $\epsilon = a/L \ll 1$. Note that the formulation of equations (\ref{eq:sbtheory})--(\ref{eq:mobility}) neglects long-ranged hydrodynamic interactions between distant parts of the filament: these interactions could be accounted for using non-local slender-body theory as in our past work on steady shear flow \citep{Liu2018}, where we found that they have a negligible effect on the dynamics.  We scale lengths by $L$, time by the characteristic relaxation time of bending deformations $\tau_r = 8 \pi \mu L^4/B$, elastic forces by the bending force scale $B/L^2$, and Brownian forces by $\sqrt{L/\ell_p}B/L^2$. The dimensionless equation of motion then reads 
\begin{equation}\label{eq:sbt}
    \partial_{t} \boldsymbol{x}(s, t)=\bar{\mu}_\text{m} \boldsymbol{u}_{\infty}(t)-\boldsymbol{\Lambda} \cdot\left[\boldsymbol{x}_{s s s s}-\left(T \boldsymbol{x}_{s}\right)_{s}+\sqrt{L / \ell_{p}} \boldsymbol{\zeta}\right],
\end{equation}
where $\boldsymbol{\zeta}$ is a Gaussian random vector with zero mean and unit variance. There are three dimensionless numbers that govern the evolution of the filament. The first one is the elastoviscous number. As mentioned in \S~\ref{sec:intro}, this serves as the measure of the effective hydrodynamic forcing and is defined as
\begin{equation}
    \bar{\mu}_\text{m} = \frac{8 \pi \mu \dot{\gamma}_m L^{4}}{B |c|}.
\end{equation}
The second one is the dimensionless time period $\rho = \dot{\gamma}_m T$, which enters the dimensionless time-periodic external flow given as $\bu_\infty(t) = \left[\sin(2 \pi \bar{\mu}_\text{m} |c| t/\rho) y,0,0\right]$. Finally, the third one is $L/\ell_p$, which captures the strength of thermal shape fluctuations. Consistent with findings from our previous work \citep{Liu2018,Chakrabarti2020}, we will see that thermal shape fluctuations have little direct effect on the filament shape dynamics, and that their primarily role is to trigger instabilities and smooth out sharp deterministic bifurcations. 

We solve equation~\eqref{eq:sbt} following the numerical algorithm outlined in \cite{Tornberg2004} and \cite{Liu2018}. For a freely suspended filament, we use the  boundary conditions: $\bx_{ss} = \bx_{sss} = \boldsymbol{0}$ and $T = 0$ at $s=\pm 1/2$. The unknown Lagrange multiplier $T(s)$ is obtained by using the constraint of inextensibility $\bx_s \cdot \bx_s = 1$. In our experiments, we have access to filament conformations in the plane of the microscope. To facilitate comparison, we thus restrict ourselves to 2D simulations where the filament is confined to the plane of the flow. 

\subsection{Conformation characterization}

To quantify the morphologies and orientations of the filaments in flow, we introduce the 2D gyration tensor~\citep{Liu2018} defined as
\begin{equation}
 G_{ij}(t) = \frac{1}{L}\int_{-L/2}^{L/2} \left [x_i(s,t) - \bar {x}_i(t) \right] \left[x_{j}(s,t) - \bar{x}_{j}(t)  \right] \md s,
\label{eq:G}
\end{equation}
where $\bar{\bx}$ is the instantaneous position of the center-of-mass of the filament. The eigenvalues of the gyration tensor ($\lambda_1$,$\lambda_2$) are combined to compute a sphericity parameter $\omega=1-4\lambda_1\lambda_2/(\lambda_1+\lambda_2)^2$, which quantifies filament deformations: it varies between $\omega\approx 1$ ($\lambda_1\gg\lambda_2\approx0$) for a straight undeformed configuration to $\omega\approx 0$ ($\lambda_1 \approx\lambda_2$) for a nearly isotropic, hence strongly deformed, shape. The eigenvector corresponding to the largest eigenvalue is used to obtain the mean orientation angle $\theta$ with respect to the negative flow direction, as illustrated in Fig.~\ref{fig:setup}{\textbf{(d)}}.

\section{Results and discussion}\label{sec:results}

\subsection{Classification of dynamics over a half period}

In an attempt to illustrate the distinct features of an oscillatory shear flow compared to its steady counterpart, let us first consider the more straightforward case of a rigid non-Brownian rod in the limit of vanishing aspect ratio $\epsilon \rightarrow 0$. A rigid rod will rotate and perform a periodic tumbling motion about its center in a steady, planar shear flow, while translating with the fluid. The tumbling motion is described by Jeffrey’s model~\citep{Jeffery1922}, which predicts that the tumbling frequency is proportional to the shear rate $\nu_t \sim\dot {\gamma}$~\citep{Jeffery1922,Tornberg2004}. 
When the rotational diffusion of the rod is taken into account, the tumbling dynamics are also affected by the additional timescale $D_r^{-1}$. As reported in previous works~\citep{Puliafito2005,Kobayashi2010,Harasim2013}, as long as $\dot\gamma \gg D_r$, thermal fluctuations only dominate over the effects of shear in a restricted angular region around the flow axis, where the effect of the noise is to drive the rod across the line $\theta=0$ in a finite time and thus initiate a new tumbling cycle. In a steady shear flow, therefore, a Brownian filament alternates between fast deterministic phases, in which it undergoes both compressional and extensional viscous stresses, and relatively long diffusive phases dominated by thermal fluctuations around the flow-aligned state. During each tumble, any deformation in the compressive quadrant of the flow is followed by relaxation in the extensional part before a new rotational cycle begins. Therefore, each tumbling phase starts with a thermally fluctuating, flow-aligned straight conformation with $\theta_0 \approx 0$.

The situation is fundamentally altered in an oscillatory flow. The mean filament orientation now oscillates with a frequency set by the flow's characteristic time period $T$. For a rigid non-Brownian rod in the plane of the flow, Jeffery's equation reads: 
\begin{equation}\label{eq:theta}
    \frac{\mathrm d \theta}{\mathrm d t} = - \sin^2 \theta(t) \sin (2 \pi t/ \rho).
\end{equation}
The orientational amplitude and frequency are thus governed by the single dimensionless number $\rho=\dot{\gamma}_m T$, which, unlike in steady shear, can be varied independently either by changing the maximum shear rate $\dot{\gamma}_m$ or the time period $T$.
In addition, the periodic flow reversal allows the filaments to explore different orientations at the beginning of each oscillation. For a fixed time period and shear rate, this initial orientation determines the relative fraction of time the filament spends in the compressive or extensile quadrants before the flow reverses at $t = T/2$. {As long as the initial filament orientation, as well as its orientation at reversal, remain sufficiently far from the flow direction, the influence of rotational diffusion should remain negligible.}  We thus expect the deformation dynamics to be sensitive to both $\rho$ and to the initial orientation $\theta_0 \equiv \theta(t=0)$ at the start of the cycle {and to be well described by Jeffery dynamics.}

\begin{figure}
  \centering
  \includegraphics[width=0.8\linewidth]{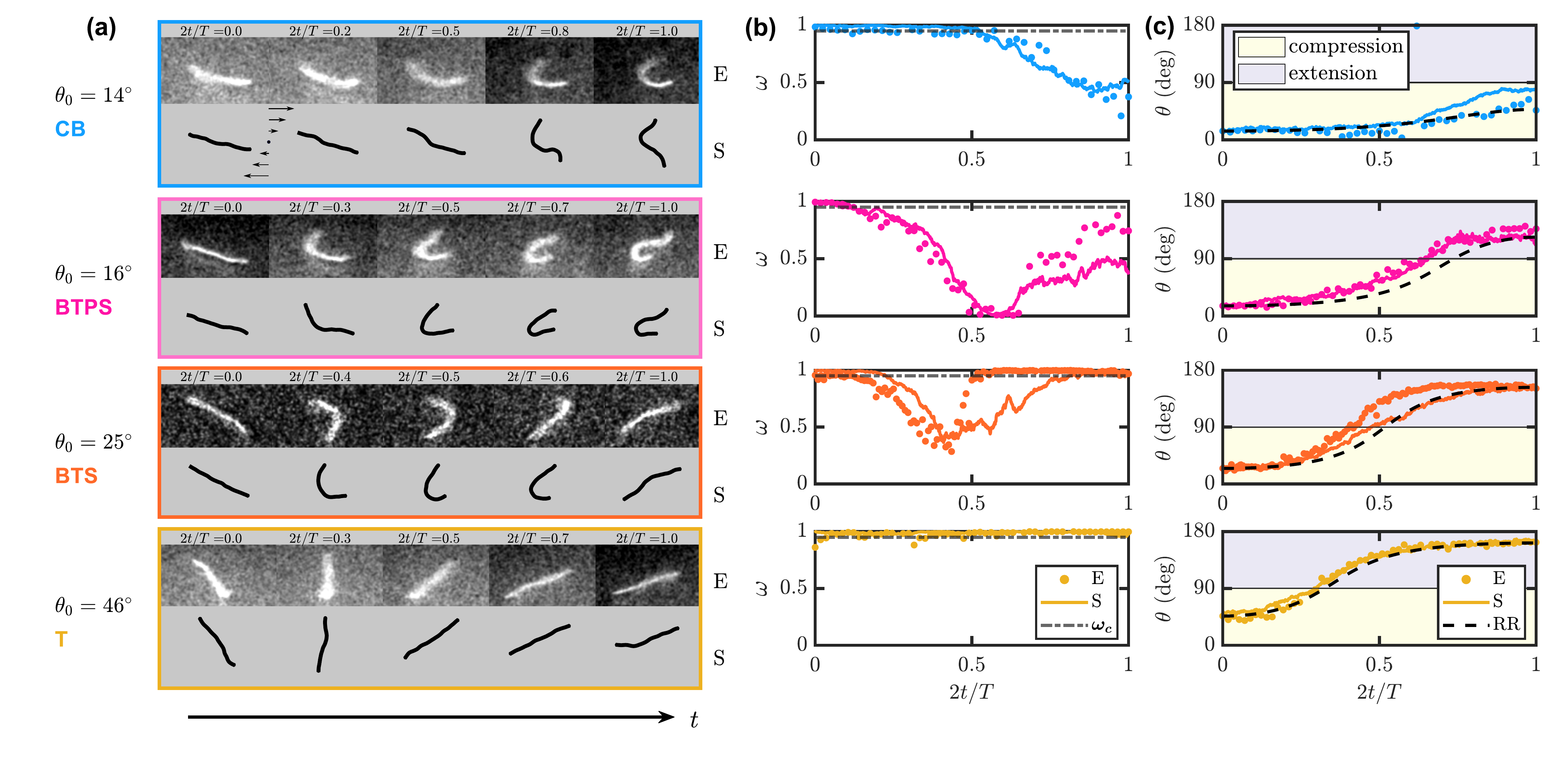}
  \caption{ The role of initial orientation $\theta_0$ on the emerging dynamics of filaments over a half time period. We classify four distinct dynamical behaviors {\bf{(a)}} by the characteristic evolution of the anisotropy parameter $\omega$ {\bf{(b)}} and orientation $\theta(t)$ defined in terms of the end-to-end vector {\bf{(c)}}. The evolution of $\theta(t)$ is further compared to a rigid non-Brownian rod model (RR, equation~(\ref{eq:theta})) as shown by the black dashed line. The symbols correspond to experiments (E) and compare well with the direct simulations (S) indicated by solid lines. Yellow/grey regions in {\bf{(c)}} corresponds  to compression and extension, respectively. In {\bf{(b)}}, $\omega_c=0.95$ is the value of $\omega$ below which a deformation is classified as a buckling event.  Parameter values: $\bar{\mu}_\text{m} = 10^4$, $\rho = 13$, and $\ell_p/L = 1.3$.}
  \label{fig:theta0}
\end{figure}

Besides, the orientational dynamics for the flexible filaments is also affected by $\bar{\mu}_\text{m}$ with the possibility of large deformations. These deformations lead to a non-reciprocal conformational evolution. As a result, the flow reversal at $t=T/2$ does not generally lead to retracing of the morphological dynamics of the first half-period, which implies that $\theta(T) \neq \theta_0$ in general. Thus, it is helpful to first restrict our analysis to the half period $0<t<T/2$, where we have precise control over all the dynamical parameters of the problem. Note that we only consider filaments that are not deformed at $t=0$. We reserve the discussion of the  filament evolution over an entire time period for \S~\ref{sec:fullperiod}.

Figure~\ref{fig:theta0} illustrates the role of the initial orientation $\theta_0$ on the emergent morphological dynamics over the half period. The mean orientation angle $\theta$ in the $x$-$y$ shear plane is measured clockwise from the negative $x$-axis; see Fig.~\ref{fig:setup}{\bf{(d)}}. For $0<t<T/2$, the shear rate is positive, which results in compressive viscous stresses for $0^{\circ}< \theta <90^{\circ}$, and extensile stresses for $90^{\circ}<\theta<180^{\circ}$. In all these examples, we fixed $\ell_p/L = 1.3$ and $\rho = \dot{\gamma}_m T = 13$. The elastoviscous number was fixed at $\bar{\mu}_\text{m} = 10^4$, which is significantly higher than the buckling threshold of $\bar{\mu}_\text{m}^c = 306.8$ for steady shear. In Fig.~\ref{fig:theta0}{\bf{(a)}}, we show snapshots of filament conformations over the half time period from both experiments and simulations, which show very good qualitative agreement. Quantitative agreement is observed in Fig.~\ref{fig:theta0}{\bf{(b)}} and {\bf{(c)}}, where for each horizontal panel we respectively plot the time evolution of $\omega$ and $\theta$ from experiments (colored symbols) and simulations (solid lines in corresponding colors). We also report in Fig.~\ref{fig:theta0}{\bf{(c)}} the evolution of the angle $\theta(t)$ expected for a non-Brownian rigid rod obeying Jeffery's equation~\eqref{eq:theta} (black dashed line). The good agreement between experiments, simulations, and the rigid-rod model indicates that Jeffery's model for orientational dynamics works well even in cases where the polymers are deformed. {This also confirms that orientational fluctuations only play a secondary role, in agreement with the fact that} we focus on relatively small values of $\rho$ (in the range of $1-20$), for which the filaments do not have enough time to get aligned with the flow axis where Brownian effects are expected to dominate. 

Depending on the initial orientation $\theta_0$ of the filament, we identify four types of dynamical behaviors, each characterized by distinct morphological dynamics over the course of the half period. Their characteristics can be summarized as follows:
\begin{itemize}

    \item \emph{ Continuous buckling (CB):} For a small initial angle close to the flow axis, $\theta_0 \sim 14^{\circ}$ (top blue panel), the filament buckles in the presence of compressive viscous loading. Here, the filament is entirely contained in the compressional quadrant over the half period, and never enters the extensional quadrant.  After initiation of buckling, it keeps getting compressed and $\omega$ reaches a minimum when $t=T/2$. 
    
    \item \emph{ Buckling then partial stretching (BTPS):} Upon slightly increasing $\theta_0 \sim 16^{\circ}$ (magenta panel), we transition to a regime where the filament rotates into the extensional quadrant of the flow, as illustrated by the angle $\theta$ in panel {\bf{(c)}}. At this point, the filament stops deforming and gets partially stretched out. This can be seen from the final conformation as well as from the parameter $\omega$.  
    
    \item \emph{ Buckling then stretching (BTS):} Upon further increase of $\theta_0 \sim 25^{\circ}$ (orange panel), we observe a dynamics similar to \emph{BTPS} with a key difference: since the filament spends less time in the compressional quadrant, its deformation is smaller when entering the extensional quadrant (at around $t=T/4$), so that it has enough time to get entirely stretched before the end of the half period. This is indicated by a distinct minimum in the evolution of $\omega$ followed by an evolution towards a straight conformation with $\omega \approx 1$.

    \item \emph{ Tumbling (T):} For a large $\theta_0 \sim 46^{\circ}$ (bottom yellow panel), the filament spends little time in the compressive quadrant of the flow. As a result, it is found to tumble without any observable deformation, in a way akin to a rigid Brownian rod. This is further quantified by the evolution of the anisotropy parameter $\omega$ that remains close to unity.
    
\end{itemize}

We emphasize that we need to distinguish buckling-induced large deformations from fluctuation-induced conformational changes in both our experiments and simulations. This becomes a challenging task for $L \sim \ell_p$. In all of the examples shown here, we have chosen a threshold of $\omega_c = 0.95$ [dash-dotted lines in Fig.~\ref{fig:theta0}{\bf{(b)}}] below which we classify a filament shape as buckled. As we will see in the following sections, this choice does not significantly alter the major conclusions of our analysis.


\begin{figure}
  \centering
  \includegraphics[width=0.8\linewidth]{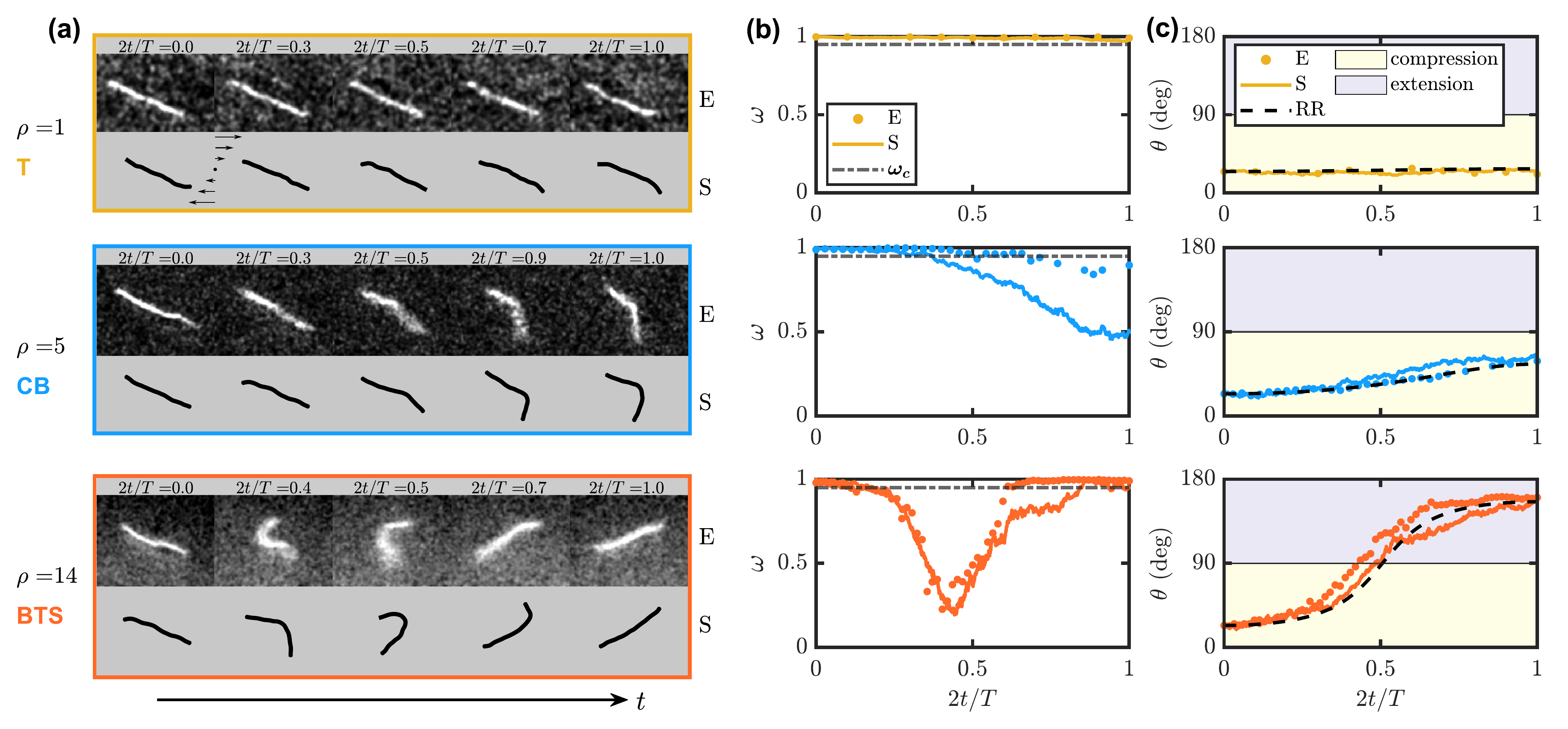}
  \caption{ The role of the dimensionless time period $\rho$ on the emerging dynamics over a half time period. The initial orientation in all these examples is fixed at $\theta_0 \approx 25^{\circ}$. Parameters: $\ell_p/L = 1.3$ and $\bar{\mu}_\text{m} = 5 \times 10^3$. {\bf{(a)}} Snapshots from experiments (E) and simulations (S) for increasing values of $\rho$. {\bf{(b)}} Evolution of the anisotropy parameter $\omega(t)$ and {\bf{(c)}} of the orientation $\theta(t)$. Color panels and legends are the same as in Fig.~\ref{fig:theta0}.}
  \label{fig:rho}
\end{figure}

Figure~\ref{fig:rho} indicates the same dynamical transitions as a function of the time period $\rho$, for a fixed $\theta_0$ and $\bar{\mu}_\text{m}$. As $\rho$ is increased, the orientation amplitude increases (panel {\bf{c}}). This results in a transition from tumbling for the smallest $\rho$ (top panel) to the occurrence of \emph{CB} (central panel) and eventually \emph{BTS} (bottom panel) dynamics for the intermediate and largest values of $\rho$, respectively. 

\begin{figure}
  \centering
  \includegraphics[width=0.35\linewidth]{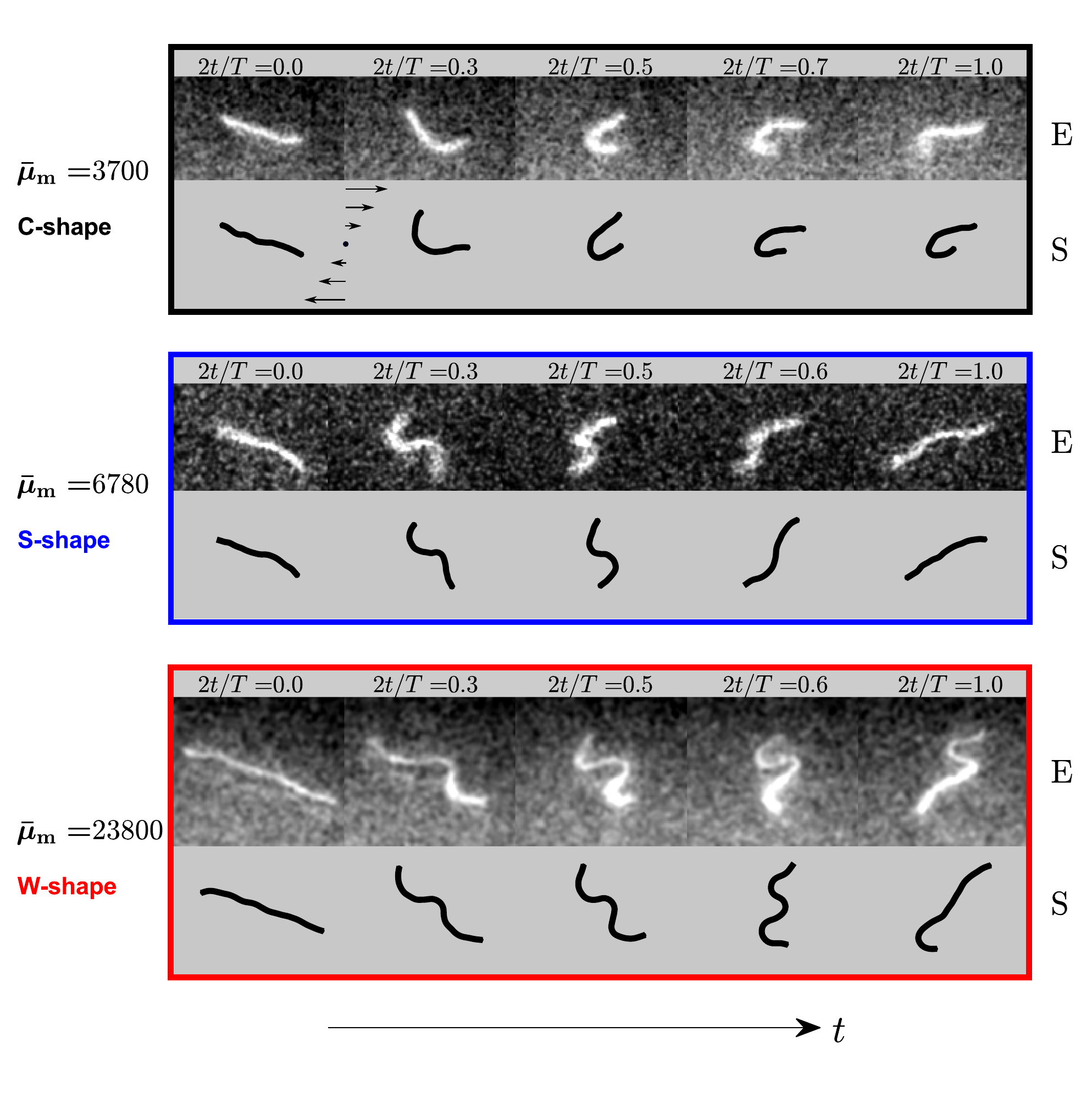}
  \caption{Various buckling modes and conformational dynamics observed with increasing values of $\bar{\mu}_\text{m}$, for a fixed initial orientation $\theta_0 \approx 20^{\circ}$ and dimensionless time period $\rho \approx 13$.}
  \label{fig:mubar}
\end{figure}

As highlighted in Figs.~\ref{fig:theta0} and~\ref{fig:rho}, the two critical aspects of an oscillatory flow are (i) the possibility of exploring many different initial orientations of the filament and (ii) of independently controlling the angular amplitude of the filament motion. In very strong flows, we can show that these two aspects beget higher-order buckling modes absent in steady shear. Figure~\ref{fig:mubar} illustrates morphological dynamics as a function of the elastoviscous number $\bar{\mu}_\text{m}$ for a fixed orientation and time period. Identical to steady shear, we find that filaments can have a characteristic $C$ shaped Euler buckling as $\bar{\mu}_\text{m}$ is increased. However, unique to the oscillatory forcing is the emergence of {$S$} and $W$ shaped configurations at large $\bar{\mu}_\text{m}$ (middle and bottom panels). These higher-order buckling modes along with global tumbling of the filament backbone are not observed in steady shear where the filaments are always aligned with the flow direction at the beginning of any time period, and as a result undergo a tank treading motion with hairpin configurations for sufficiently large $\bar{\mu}_\text{m}$ \citep{Liu2018}.


\subsection{Suppression of buckling instabilities in strong flows} \label{sec:buckling}

A remarkable feature in Fig.~\ref{fig:rho} is the absence of buckling instabilities at small $\rho$ (top yellow panel), even though the filament is entirely contained in the compressional quadrant of the flow and the elastoviscous number exceeds the theoretical buckling threshold, $\bar{\mu}_\text{m} \gg \bar{\mu}_\text{m}^c$. A linear stability analysis in the spirit of \cite{Becker2001} would suggest the emergence of deformed conformations. However, in the presence of a high-frequency periodic forcing we observe suppression of finite-size deformations. Interestingly, as illustrated in the bottom panel of Fig.~\ref{fig:theta0}, filament buckling is absent even at large $\rho$ for initial orientations close to $45^{\circ}$, for which the compression rate is maximum.

These two examples highlight the subtle interplay of two nonlinear effects beyond the predictions of a linear stability theory. First, filament orientational dynamics dictate the duration and strength of the effective compressive viscous loading acting along the filament backbone, {depending on both filament orientation and instantaneous flow strength}. This is sensitive to the initial orientation $\theta_0$ and the time period $\rho$. Second is the coupling between the {duration of filament exposure to compressive forces}  and the characteristic time scale over which deformations grow to be detectable over thermal fluctuations. In \S~\ref{sec:landau}, we will develop a weakly nonlinear Landau theory that accounts for all of these effects, but first, we focus on their quantitative characterization from experiments and simulations.

\begin{figure}
	\centering
	\includegraphics[width=0.8\linewidth]{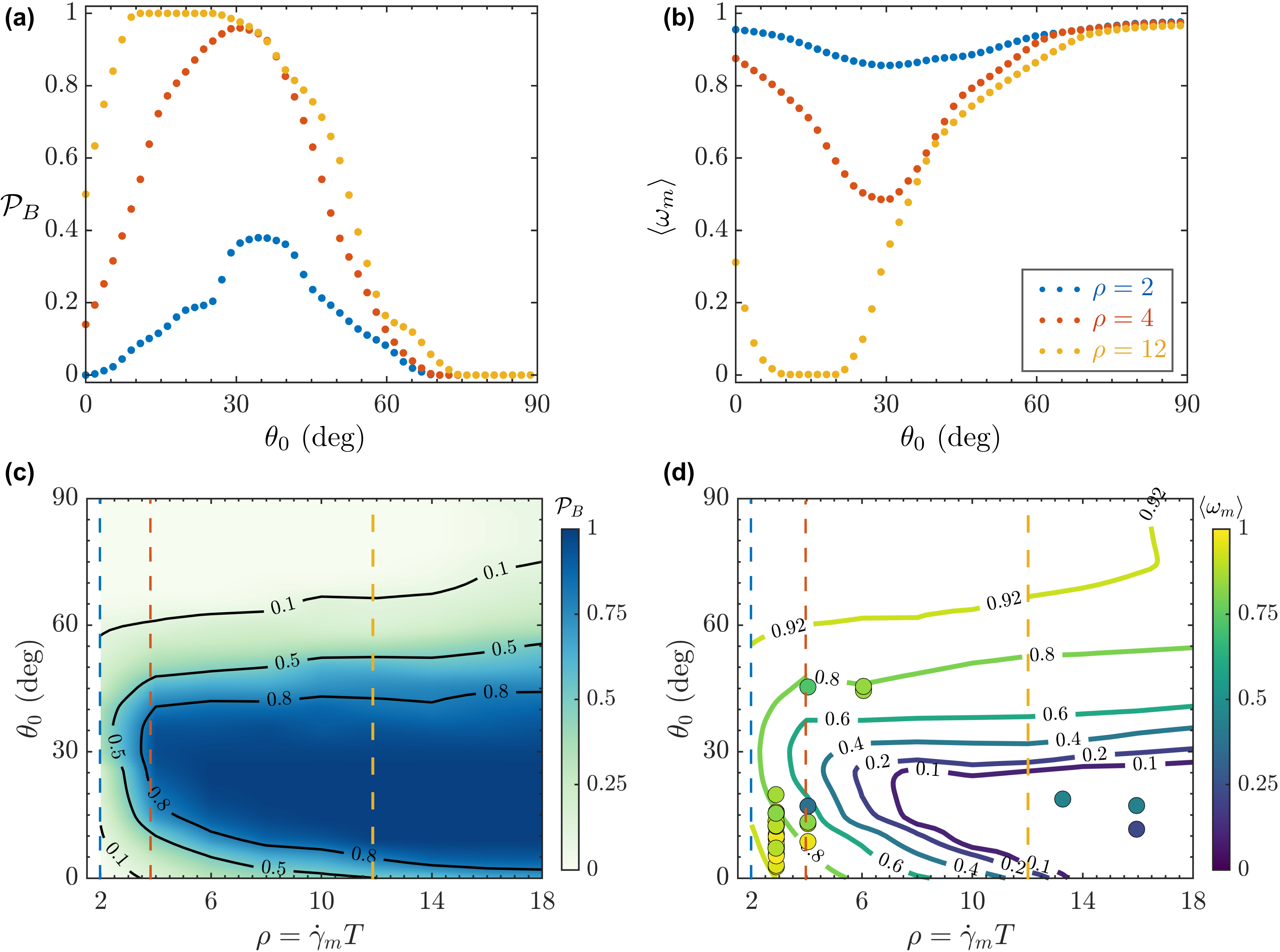}
	\caption{{\bf{(a)}} Buckling probability $\mathcal{P}_B$ as a function of $\theta_0$ as obtained from simulations (50 independent runs) for three different dimensionless time periods $\rho$. {\bf{(b)}} The ensemble averaged minimum value $\omega_m$ of the deformation parameter, corresponding to the maximum deformation, from the same simulations. We show in panel {\bf{(c)}} and {\bf{(d)}} the phase charts of $\mathcal{P}_B$ and $\omega_{m}$ versus $(\rho,\theta_0)$ for a larger set of dimensionless periods $\rho\in[2,18]$. In {\bf{(d)}}, the maximum deformation from simulations (isolines) is also compared to experimental data (symbols) under identical conditions. The vertical colored lines in {\bf{(c)}} and {\bf{(d)}} indicate the constant $\rho$ slices shown in {\bf{(a)}} and {\bf{(b)}}. Parameter values: $\bar{\mu}_\text{m} = 2\times10^4$, $\ell_p/L = 1.3$. }
	\label{fig:pbcross}
\end{figure}

To this end, we performed simulations by systematically varying $\rho \in [2, \ 18]$ and $\theta_0 \in [0, \ \pi/2)$ while keeping $\bar{\mu}_\text{m} = 2\times 10^4 \gg \bar{\mu}_\text{m}^c$ and $\ell_p/L = 1.3$ fixed. For each combination of $(\rho,\theta_0)$, we performed 50 numerical simulations with different random seeds over one half period for statistical averaging. This then allows us to define and estimate a probability of buckling as
\begin{equation}
    \mathcal{P}_B(\rho,\theta_0) = \frac{\mathcal{N}(\omega(t) \le \omega_c)}{\mathcal{N}_{\text{tot}}}\biggr|_{(\bar{\mu}_\text{m},\ell_p/L)},
\end{equation}
where $\mathcal{N}(\omega(t) \le \omega_c)$ is the number of cases for which the filament buckled at some point during the half period and $\mathcal{N}_{\text{tot}} = 50$.

{Figure~\ref{fig:pbcross}{\bf{(a)}} displays the buckling probability as a function of the initial filament orientation $\theta_0$ for three different values of the dimensionless time period $\rho$. For $\theta_0 \gtrsim 70^{\circ}$, instability growth is not sufficient to lead to detectable deformations and, irrespective of $\rho$, the buckling probability $\mathcal{P}_B$ is small. For intermediate initial angles, buckling probabilities are significant, and we note that for a given $\theta_0$, the probability of buckling $\mathcal{P}_B$ increases with increasing $\rho$. When initial angles get close to $\theta_0 \sim 0^{\circ}$ a decrease in the buckling probability is again observed.}

Figure~\ref{fig:pbcross}{\bf{(b)}} shows the ensemble-averaged minimum of $\omega(t)$ reached during the evolution over the half period and serves as a measure of the deformation. The existence of strongly deformed filament conformations with a small value of $\langle \omega_m \rangle$ correlates with a probability of buckling close to unity. These behaviors are further corroborated in the phase plots shown in Fig.~\ref{fig:pbcross}{\bf{(c)}} and {\bf{(d)}}, where we systematically varied the dimensionless period in the range of $\rho=2-18$. In panel {\bf{(d)}}, superimposed on the simulations, we also indicate experimental results under identical flow conditions and find good quantitative agreement. 

{These observations are in qualitative agreement with the nonlinear effects discussed above. For initial orientations larger than $\theta_0 \gtrsim 70^{\circ}$, the time spent by the filament in the compressional quadrant is not sufficient for the instability to grow, regardless of the value of $\rho$. For small initial angles, the growth of the instability is limited by the combined effect of the reduced rotational excursion for these initial angles, which impedes the filament from reaching maximum compression, and of the time-varying flow. }

\subsection{Weakly nonlinear Landau theory \label{sec:landau}}

In the previous section, we highlighted that even in strong flows, filaments may not undergo detectable buckling instabilities in contrast to linear stability theory. To explain this finding quantitatively, we propose here a weakly nonlinear Landau theory for filament deformations. 

Our theory neglects Brownian fluctuations, and we scale time with the inverse shear rate $\dot{\gamma}_m^{-1}$. In this case, the non-Brownian SBT equation becomes
\begin{equation}\label{eq:sbtnonbr}
\bar{\mu}_\text{m} \partial_{t} \boldsymbol{x}(s, t)=\bar{\mu}_\text{m} \boldsymbol{u}_{\infty}(t)-\boldsymbol{\Lambda} \cdot\left[\boldsymbol{x}_{s s s s}-\left(T \boldsymbol{x}_{s}\right)_{s}\right],
\end{equation}
where $\bu_\infty(t) = \left[\sin(2 \pi t/\rho) y, 0, 0 \right]$. We start our analysis by recapitulating the linear stability analysis \citep{Becker2001}.
The base state of the filament is an undeformed straight rod. Suppose that at any instant the rod makes an angle $\theta(t)$ with the negative $x$ axis. We describe the instantaneous orientation of the rod by the director $\bp(t) = -\cos \theta \uvc{x} + \sin \theta \uvc{y}$. We also define the vector orthogonal to the director $\bp^\perp(t) = \sin \theta \uvc{x} + \cos \theta \uvc{y}$.  The orientation of the rod is assumed to evolve according to Jeffery's equation~\eqref{eq:theta}. 
Associated with this is a time-varying parabolic tension profile given by 
\begin{align}
    T_0(s,t) = \frac{\bar{\mu}_{\text{inst}}(t)}{8}\left(s^2 - \frac{1}{4}\right), 
    \end{align}
    where
\begin{align}
    \bar{\mu}_{\text{inst}}(t) = \bar{\mu}_\text{m} \sin (2 \pi t/ \rho) \sin (2 \theta)
\end{align}
is the instantaneous elastoviscous number experienced by the filament as it rotates in the unsteady flow. In order to first understand the linear stability, we perturb the filament in the transverse direction. The perturbed state is given by $\bx = s \bp + h_\perp(s,t) \bp^\perp$, where $h_\perp(s,t) \sim \mathcal{O}(\varepsilon)$ with $\varepsilon \ll 1$. Linearizing around the base state results in a linear equation for $h_\perp(s,t)$ given by $ \bar{\mu}_\text{m} \partial_t h_\perp = \bar{\mu}_{\text{inst}}(t)\mathcal{L}[ h_\perp]/2  - \partial_s^4 h_\perp$, where $\mathcal{L}$ is the differential operator
\begin{equation}
    \mathcal{L}[f] = \left[1 + s \partial_s + \frac{1}{4}\left(s^2-\frac{1}{4}\right) \partial_s^2 \right]f(s). 
\end{equation}
Following \cite{Becker2001}, we use the ansatz of normal modes and write $h(s,t) = \phi(s) \exp(\sigma t)$, where $\sigma$ is the growth rate. This leads to an eigenvalue problem for the growth rate $\sigma$ and the eigenfunction $\phi(s)$, from which we find that there is an Euler buckling instability when the elastoviscous number exceeds $\bar{\mu}_\text{m}^c \approx 306.8$. At the onset of this instability, we have $\sigma = 0$ and $\phi = \phi^c(s)$. This means that the eigenfunction at the critical threshold satisfies 
\begin{equation}\label{eq:linrel}
    \mathcal{L}[ \phi^c] = \frac{2}{\bar{\mu}_\text{m}^c} \partial_s^4 \phi^c.
\end{equation}

We now proceed to describe the nonlinear dynamics of the filament away from the instability threshold. For that purpose, we develop a weakly nonlinear theory in the vicinity of the first bifurcation. We expand the transverse perturbation on the basis of the first mode of deformation $\phi^c(s)$ as $h_\perp(s,t) = A(t) \phi^c(s)$, where $A(t) \sim \mathcal{O}(\varepsilon)$. Transverse perturbations induce changes in length along the axis of the filament. As we will see, these length changes are higher order in $A(t)$. We account for this by introducing axial perturbations $h_\pll(s,t)$. The perturbed filament conformation can then be written as:
\begin{equation}
    \bx(s,t) = \left(s + h_\parallel\right) \bp + h_\perp \bp^\perp.
\end{equation}
Similarly, we perturb the tension around its base state $T_0(s,t)$ as:
\begin{equation}
    T(s,t) = T_0(s,t) + \tilde{T}(s,t).
\end{equation}
We choose to describe the filament backbone and its mean orientation with the rigid rod model, which is supported by the observations of Fig.~\ref{fig:theta0}. So at any instant, we assume that the orientation $\theta(t)$ follows equation~\eqref{eq:theta}. We next outline the set of steps to obtain the evolution equation for the amplitude $A(t)$:
\begin{itemize}
    \item \textbf{ Inextensibility:} Filament inextensibility implies that $\bx_s \cdot \bx_s = 1$, which yields:
    \begin{equation}\label{eq:inexten}
        \bx_s \cdot \bx_s = 1 + \left[2 \partial_s h_\pll + (\partial_s h_\perp)^2 \right] + (\partial_s h_\pll)^2.
    \end{equation}
    We first note that in absence of any axial perturbations the inextensibility condition was satisfied up to $\mathcal{O}(\varepsilon^2)$. This points to the fact that $h_\pll \sim \mathcal{O}(\varepsilon^2)$. We can now satisfy inextensibility up to $\mathcal{O}(\varepsilon^4)$ and obtain $h_\pll(s,t)$ by setting the second term in equation~\eqref{eq:inexten} to zero. Thus we obtain
    \begin{equation}
        h_\parallel(s,t) = -\frac{1}{2}\int_{-1/2}^s \left[\partial_{s'} h_\perp(s',t) \right]^2 \mathrm{d} s'= {A(t)^2} v(s),
    \end{equation}
    where we have defined $v(s) = -\frac{1}{2}\int_{-1/2}^{s} \phi^c_{s'}(s')^2 \md s'$.
    
    \item \textbf{ Tension perturbation:} In order to solve for the tension, we first derive an auxiliary equation from the SBT using the fact that $\partial_t \bx_s \cdot \bx_s = 0$. This provides a second-order ODE for the tension $T(s,t)$ as:
    \begin{equation}
        2 T_{s s}- (\bx_{ss}\cdot \bx_{ss}) T=-\bar{\mu}_\text{m} \bx_s \cdot \partial_s \bu -6 \bx_{s s s} \cdot \bx_{s s s}-7 \bx_{s s} \cdot \bx_{s s s s} \equiv \mathcal{R}.
    \end{equation}
    Retaining terms up to $A(t)^3$, the equation for the tension perturbation becomes:
    \begin{align}
    2 \tilde{T}_{ss}  =  &\left(\bar{\mu}_\text{m} A(t) \cos(2 \theta) \phi^c_s + A(t)^2 \left[-\bar{\mu}_\text{m} \sin(2 \theta) (\phi^c_s)^2 - 6 (\phi^c_{sss})^2 - 7 \phi^c_{ss} \phi^c_{ssss}  \right] \right) \sin(2 \pi t/ \rho) \nonumber  \\
    &\hspace{2cm}+ A(t)^2 (\phi^c_{ss})^2 T_0(s,t).
    \end{align}
    The above linear system can be solved numerically along with the boundary condition $\tilde{T}(s=\pm 1/2) = 0$.
    
    \item \textbf{ Amplitude equation at the linear order:} Before deriving the nonlinear amplitude equation, we first consider the nature of the linearized amplitude equation.  At the linear order we have: 
    \begin{equation}
        \bar{\mu}_\text{m} \frac{\md A}{\md t} \phi^c(s) = \left(\bar{\mu}_{\text{inst}}(t) \mathcal{L}[\phi^c]/2 - \partial^4_s \phi^c \right)A(t).
    \end{equation}
    On using equation~\eqref{eq:linrel}, we can simplify the above equation and obtain:
    \begin{equation}
        \frac{\md A}{\md t} = \big\langle \phi^c,\partial^4_s \phi^c \big\rangle \left[\frac{\bar{\mu}_{\text{inst}}(t) - \bar{\mu}_\text{m}^c}{\bar{\mu}_\text{m}\bar{\mu}_\text{m}^c} \right] A(t),
    \end{equation}
    where $\langle f,g \rangle$ is the inner product of two functions. The above relation amounts to recasting the results of the linear stability analysis, and predicts exponential growth of the amplitude $A(t)$ whenever $\bar{\mu}_{\text{inst}} > \bar{\mu}_\text{m}^c$.
    
    \item \textbf{ Landau equation for the amplitude:} In order to derive a nonlinear amplitude equation, we make use once again of the SBT equation. Using our ansatz for the perturbations, we project the equations onto the transverse direction. Retaining all the terms up to order $A(t)^3$ and using the solution from the perturbed tension equation, we can obtain a nonlinear amplitude equation of the form: 
    \begin{equation}\label{eq:landau}
    \frac{\md A}{\md t} = \big\langle \phi^c,\partial^4_s \phi^c \big\rangle \left[\frac{\bar{\mu}_{\text{inst}}(t) - \bar{\mu}_\text{m}^c}{\bar{\mu}_\text{m}\bar{\mu}_\text{m}^c} \right] A(t) + \frac{1}{\bar{\mu}_\text{m}}\big \langle \phi^c, {\Omega} \big \rangle.
    \end{equation}
    In the above equation, $\Omega$ captures the effect of nonlinearities and is given by:
    \begin{equation}
        {\Omega} = A(t)^2 v(s) \sin^2 \theta + A(t) \left[2 \phi^c_s \tilde{T}_s + \tilde{T} \phi^c_{ss} \right] - A(t)^3 \left[\phi^c_{s} v_{ssss} + (\phi^c_s)^2 \phi^c_{ssss} \right].
    \end{equation}
    
    \item \textbf{ Numerical solution:} We integrate equation~\eqref{eq:landau} numerically using a fourth order Runge-Kutta scheme. The amplitude equation is complemented with equation~\eqref{eq:theta}, which provides the instantaneous orientation of the filament. At every time step, we solve the linear system for the tension perturbation $\tilde{T}(s)$ to compute the nonlinear terms ${\Omega}$. 
\end{itemize}

\begin{figure}
  \centering
  \includegraphics[width=0.8\linewidth]{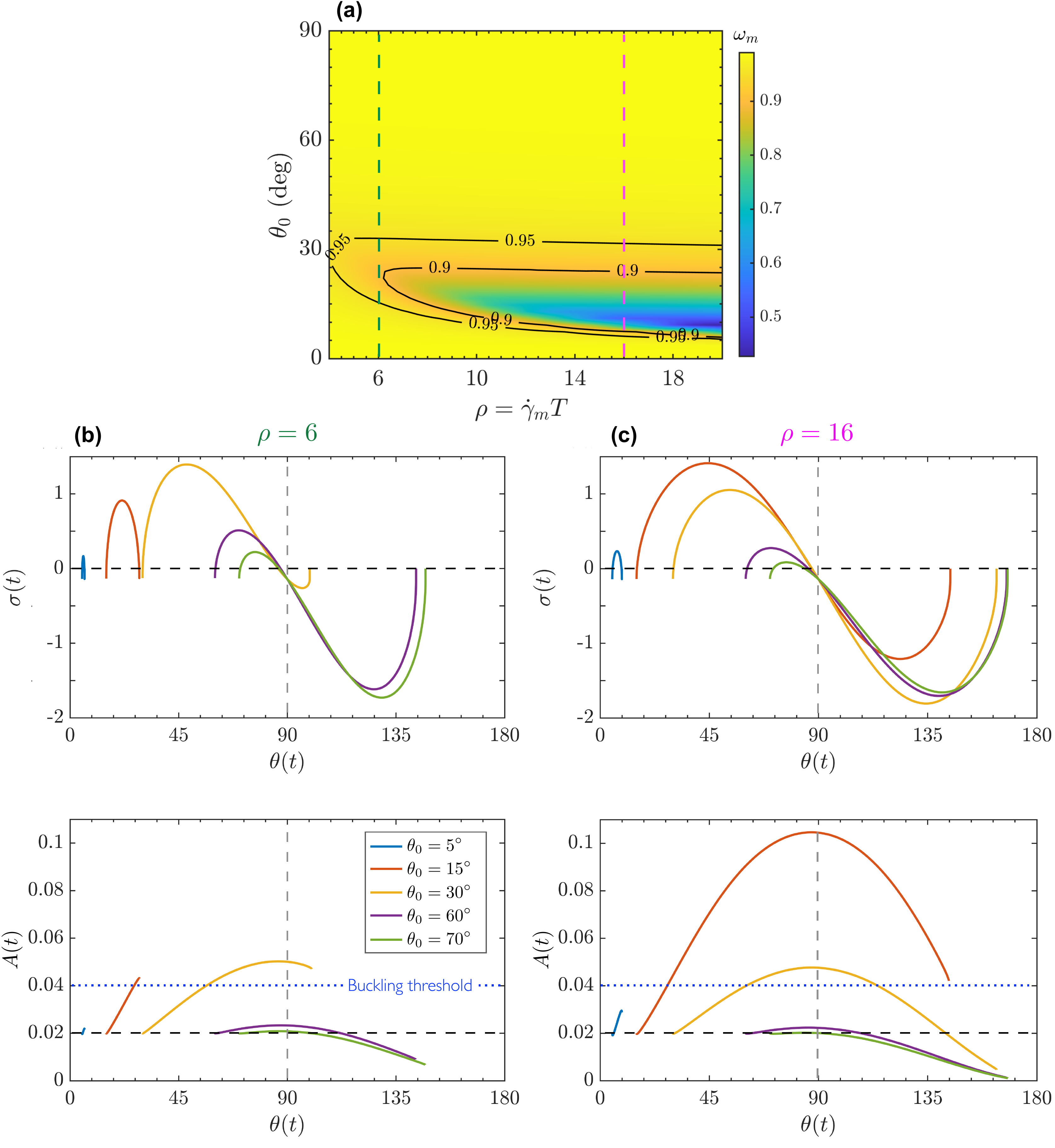}
  \caption{{\bf{(a)}} Phase plot of the maximum deformation for $\bar{\mu}_\text{m} = 3.5 \times 10^3$ as computed from the nonlinear Landau theory outlined in \S~\ref{sec:landau}. The two vertical colored lines indicate slices of $\rho = 6$ [panel {\bf{(b)}}] and $\rho = 16$ [panel {\bf{(c)}}], respectively, for which we show the evolution of the growth rate, $\sigma(t)$, and the perturbation amplitude, $A(t)$, as  functions of the instantaneous orientation of the filament, for several values of the initial angle $\theta_0$.}
  \label{fig:eigtime}
\end{figure}

In our numerical explorations, we varied $\rho$ and $\theta_0$ and studied the evolution of filament deformation predicted from the nonlinear theory. We initiated the simulations by perturbing the filament backbone as $\bx(s,t=0) = s \bp + A_0 \phi^c(s) \bp^\perp$, where $A_0$ is the initial perturbation amplitude. For all our simulations we chose $A_0 = 0.02$ for the initial amplitude, which is the typical magnitude of transverse fluctuations of a freely suspended actin filament with $\ell_p/L \sim \mathcal{O}(1)$ as considered in our experiments. We characterize an event as buckling in our theory whenever the amplitude exceeds $A \approx 0.04$, which corresponds to an anisotropy parameter $\omega_c \approx 0.95$. We show the phase-chart of minimum $\omega_m$ computed from the theory in Fig.~\ref{fig:eigtime}{\bf{(a)}}. It predicts that buckling is favored only for certain ranges of $\theta_0$ and that its probability increases with increasing $\rho$, an observation consistent with our experiments and Brownian simulations shown in \S~\ref{sec:buckling}. 

In order to explain this behavior, we now plot the evolution of the growth rate $\sigma(t)$ and the amplitude of the perturbation $A(t)$ as functions of the orientation angle of the filament $\theta(t)$ from the nonlinear model. Typical data for two values of $\rho$ and several initial orientations $\theta_0$ are reported respectively in panels {\bf{(b)}} ($\rho=6$) and {\bf{(c)}} ($\rho=16$) of Fig.~\ref{fig:eigtime}.

A number of observations can be made from these figures. First, the range of orientation angles traversed by the filament varies strongly as a function of the initial orientation, even for a fixed value of $\rho$ (compare for example blue and red curves on Fig.~\ref{fig:eigtime}b). This is linked to the fact that the Jeffery dynamics are slow close to alignment with the negative flow direction, but faster for increasing initial angles. Filaments starting aligned close to the negative flow direction thus do not rotate very far compared to filaments with initial angles closer to 45$^\circ$. Second, filaments that start close to 90$^\circ$ will spend little time in the compressive quadrant, where positive growth rates can exist. In a time independent shear flow, all filaments will experience maximum compression and thus a maximum growth rate at 45$^\circ$. In the case of an oscillatory flow, the flow forcing itself is time-dependent and the maximum growth rate results from a combination of the orientation angle and the time-dependent flow intensity, as is clear from the figures where maximum growth rates are not necessarily observed at 45$^\circ$, even for filaments sweeping by this orientation. Comparing panels {\bf{(b)}} ($\rho=6$) and {\bf{(c)}} ($\rho=16$) of Fig.~\ref{fig:eigtime}, it is obvious that larger values of $\rho$ lead to larger growth rates over more extended ranges of angles. It is also interesting to note that negative growth rates are observed even in the compressive quadrant, for situations where $\bar{\mu}_\mathrm{inst}$ remains below the buckling threshold.

{As pointed out above, to observe a buckling instability it is not sufficient for the filament to experience a positive growth rate, but the perturbation also needs sufficient time to grow so as to lead to a measurable signature in the anisotropy parameter $\omega$. This can be seen from the bottom row of Fig.~\ref{fig:eigtime}, that shows the instantaneous amplitude resulting from the cumulated growth of the perturbation during rotation of the filament through the compressive quadrant.} These examples thus underscore a key aspect of our experimental measurements: linear instability does not always correspond to detectable deformations of the filament. 

This critical distinction between linear instability and measurable deformations also applies to steady shear flows, even if to a lower extent, but seems to have been unreported. We further elaborate on this point in Appendix~\ref{app:B}.


\subsection{Dynamics over one full period \label{sec:fullperiod}}

\begin{figure}
  \centering
  \includegraphics[width=0.8\linewidth]{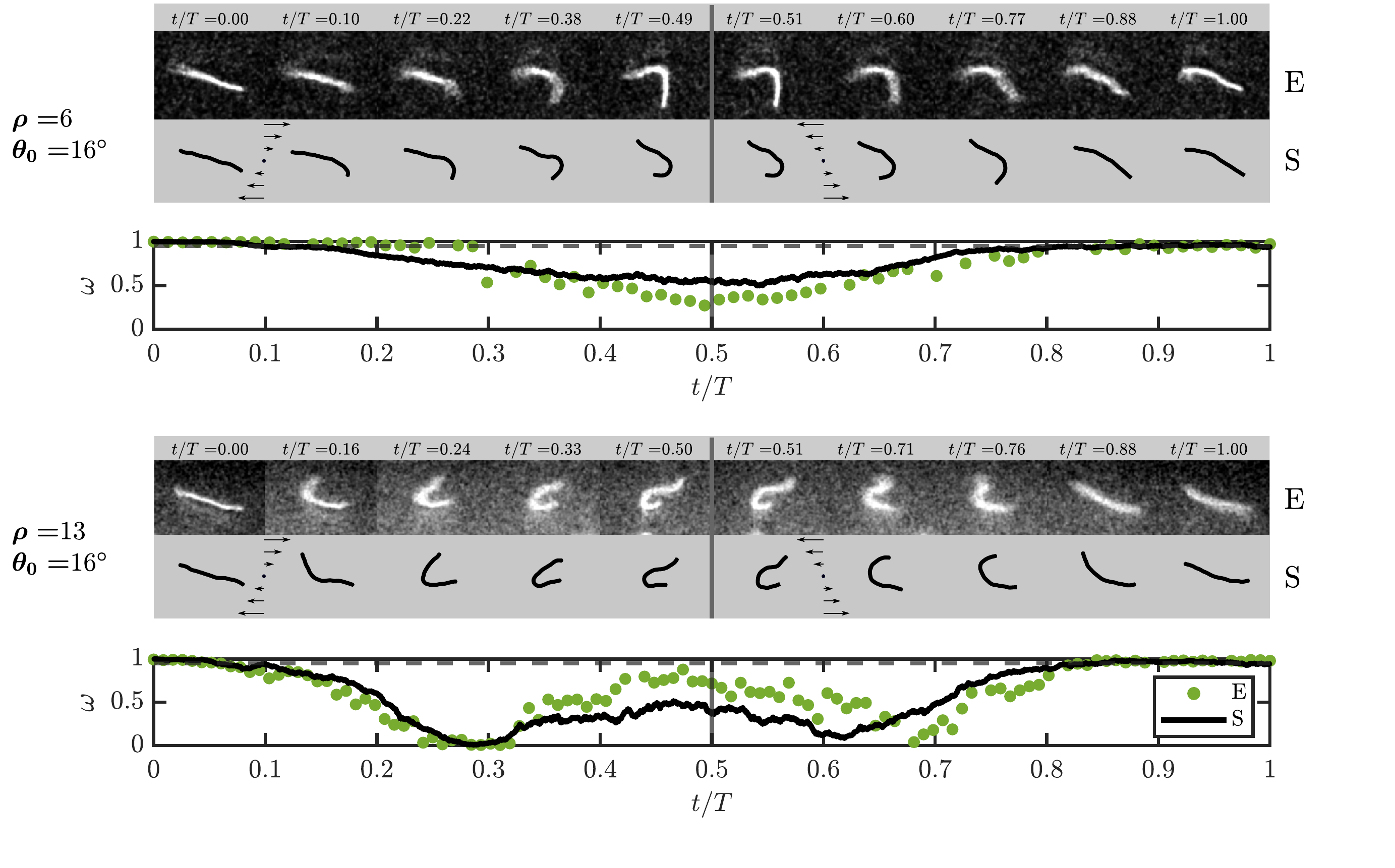}
  \caption{Dynamical features over a complete period of tumbling. During the second half period, the filament can either be continuously stretched by the flow (top) or compressed once again (bottom). Parameter values: $\ell_p/L = 1.3$, $\bar{\mu} = 1 \times 10^4$.}
  \label{fig:fullper}
\end{figure}

So far, we have focused on the deformation dynamics in the first half of the forcing period, considering only the particular case of a straight filament at $t=0$. This has helped us understand the role of the time period $\rho$ and the initial orientation $\theta_0$ that can both be controlled independently and can be systematically varied. We now discuss briefly
the filament dynamics over one entire period of the oscillatory flow. The key aspect here is that the orientation and the deformation of the filament at the beginning of the second half of the period is governed by its deformation and orientational dynamics during the first half, giving rise to new interesting dynamics. In particular, one also has to consider cases in which the filament is already deformed at the beginning of the second half cycle. The interplay between oscillation period, flow strength, initial orientation and deformation of the filament leads to a very rich spectrum of dynamical modes. Typical examples are reported in Fig.~\ref{fig:fullper}. In the top panel, the filament undergoes a continuous buckling (\emph{CB}) event during the first half cycle. The filament remains oriented in the extensional quadrant when the flow is reversed. As a result, it is stretched by the flow during the second half cycle, and the anisotropy parameter increases continuously. As $\rho$ is increased (bottom panel), the dynamics in the first half of the period transitions to a \emph{BTPS} event. Consequently, when the shear is reversed, the filament first experiences compression, deforming  from an already buckled state. As the filament enters the extensional quadrant, it is then stretched out by the flow. 

These examples give a glimpse of the complex evolution of the filament deformation over multiple cycles. Since the deformation in general brings about irreversible orientational dynamics ($\theta(T) \neq \theta_0$), the asymptotic filament dynamics is expected to be characterized by the emergence of a chaotic behavior, even in the athermal limit {and at small Reynolds number}.  


\section{Conclusion}\label{sec:concl}
 Using stabilized actin filaments as a model system, we have systematically explored the dynamics of semiflexible polymers in oscillatory shear flows. Our microfluidic experiments were shown to be in excellent agreement with Brownian simulations based on Euler-Bernoulli elasticity and local slender-body theory hydrodynamics, and were also well described by a weakly nonlinear Landau model. Through a systematic variation of the filament length and of the shear rate and frequency of the imposed flow, we characterized the orientational and morphological dynamics of the filaments over a half oscillation period of the flow and identified four distinct modes of deformation that are unique to periodic forcing. 

For a given flow strength (elastoviscous number $\bar{\mu}_\mathrm{m}$), these morphological dynamics were shown to be sensitive to the dimensionless time period $\rho$ of the flow and initial orientation $\theta_0$ of the filament, which can take on any value at the start of an oscillation cycle. These two variables govern the sign, magnitude and duration of the viscous stresses experienced by the filament as it rotates through the applied flow, giving rise to a rich variety of behaviors from rigid tumbles to elastoviscous buckling followed with varying levels of stretching. In the case of buckling, a wide range of deformations modes ($C$, $S$, and $W$ modes) was also found to emerge. These findings are in contrast with the case of steady shear, where every filament tumble initiates from the same flow-aligned stretched conformation, resulting in a more constrained set of deformation modes ($C$ modes and tank-treading dynamics) that are entirely governed by the constant applied flow strength \citep{Liu2018}. 

Another striking feature of oscillatory shear flow is the absence of observable buckling in strong flows at high oscillation frequencies (low values of $\rho$), even when the maximum elastoviscous number greatly surpasses the buckling threshold for steady shear ($\bar{\mu}_\mathrm{m}\gg\bar{\mu}_\mathrm{m}^c$). This is a consequence of the short duration of time during which compressive stresses are applied before the flow changes direction: the half period is too short for flow-induced deformations to grow sufficiently to be detected over thermal shape fluctuations, even when buckling is predicted to occur. Our simulations were used to calculate a probability of buckling, which peaks at a value close to one for $\theta_0\approx 25^\circ$ in strong flows, but drops sharply when $\rho\lesssim O(1)$. 

Our study has focused on characterizing the dynamics over a half period of the flow, and only briefly addressed the case of a full period in \S \ref{sec:fullperiod}. There, the dynamics is even richer as the conformation of the filament at the end of the first half period may involve significant deformation. The evolution of filament conformations over multiple time periods remains an open question of great interest: for instance, it is not known whether the filament dynamics would cycle, perhaps chaotically, through a variety of morphological transitions, or if the system may end up selecting specific deformation modes corresponding to preferred initial orientations after multiple oscillations. Some of the findings from our study also hint at interesting rheological behaviors for suspensions of semiflexible polymers in time-periodic flows. In steady shear, our previous work demonstrated that the onset of buckling instabilities is accompanied by enhanced shear-thinning and an increase in normal stress differences \citep{Chakrabarti2021a}. In an oscillatory flow, we expect the same effect to occur but also anticipate a non-trivial dependence on the flow period, with a possible increase in viscosity at high frequencies due to the suppression of buckling in that regime. Some of these outstanding questions will be addressed in future work.

\begin{acknowledgments}

We are grateful to the team of Guillaume Romet-Lemonne and Antoine Jégou for providing purified actin and to Michael Shelley for fruitful discussions. A.L. and F.B. acknowledge support from the ERC Consolidator Grant PaDyFlow under grant agreement 682367. We thank Institut Pierre-Gilles de Gennes (Investissements d'avenir ANR-10-EQPX-34).

\end{acknowledgments}

\appendix 

\section{Velocity profile and shear rate in the observation plane} \label{app:A}

In this section we show that the flow profile in the channel horizontal plane is always parabolic over the course of a typical oscillatory experiment, and highlight the procedure used to obtain the time-dependent, local shear rate experienced by the filament.

\begin{figure} 
  \centering
  \includegraphics[width=0.8\linewidth]{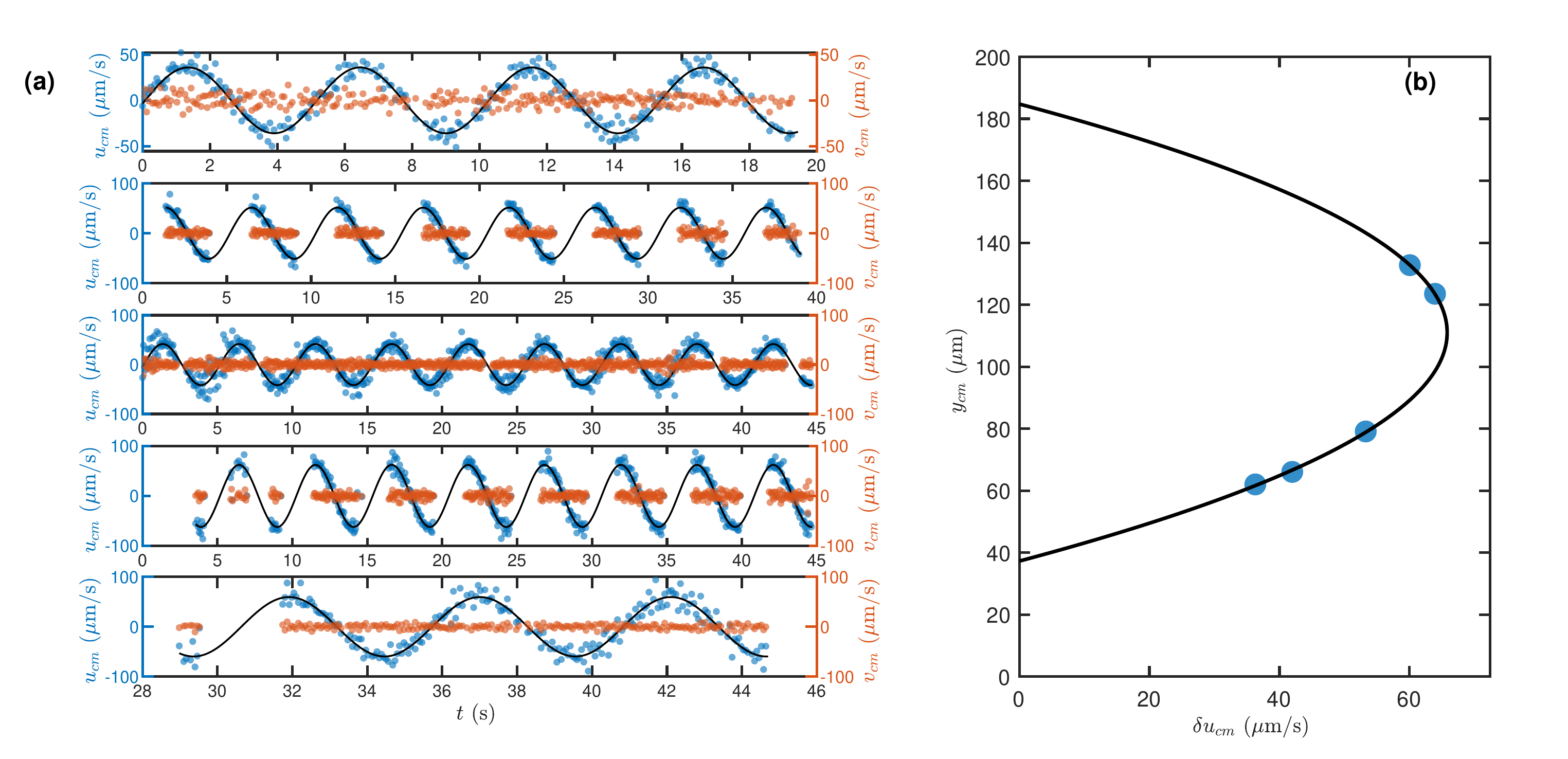}
  \caption{\textbf{(a)} Parallel ($u_{cm}$) and perpendicular ($v_{cm}$) center-of-mass velocities of filaments located in different regions of the observation plane. \textbf{(b)} The velocity amplitude $\delta u_{cm}$ vs. $y$-position of the filament is parabolic (solid line), which attests a Poiseuille-like flow profile.}
  \label{fig:vel}
\end{figure}

Figure~\ref{fig:vel}{\bf{(a)}} shows the measured velocities perpendicular ($v_{cm}$) and parallel ($u_{cm}$) to the flow of $n=5$ filaments that have been tracked during a typical oscillatory experiment with $T=$ \SI{5}{s}. The intermittent lack of experimental points in the data of some filaments is due to the fact that these filaments are located close to the borders of the image, and so temporarily lost during the oscillation. 
First, we note that all the measured transverse velocities only fluctuate around zero. This indicates the absence of any significant drift perpendicular to the flow. As expected, periodic oscillations are observed for the velocities in the streamwise direction. Fitting to a sinusoidal function (solid black lines) yields, for each filament, an independent estimate of the period $T$, phase $\phi$, and amplitude $\delta u_{cm}$ of the oscillation. We get $T=5.094\pm0.003$ \SI{}{s} and $\phi=(-0.09\pm0.02)$ \SI{}{rad} (mean $\pm$ std) for the period and phase, respectively. Since the camera and the pressure controllers are synchronized, $\phi$ corresponds to the actual phase shift between the imposed pressure and the flow. It indicates that the flow only slightly lags behind the input pressure. Figure~\ref{fig:vel}{\bf{(b)}} shows the velocity amplitude $\delta u_{cm}$ inferred from the fit as a function of the $y$-position of the filament $y_{cm}$. As seen, the profile is well described by a Poiseuille flow (black solid line). The fit also yields a channel width of $W=147.5\pm 16$ \SI{}{\micro m}, in good agreement with the expected one. As the flow is parabolic at any time, and considering the flow gradient as constant over the length scale of the filament (\S~\ref{sec:m&m}), the instantaneous shear rate is computed as:
\begin{equation}
\dot\gamma(t) = \frac{u_{cm}(t)\left[W-2\Delta y_{cm}\right]}{\Delta y_{cm}\left[W-\Delta y_{cm}\right]},
\end{equation}
where $\Delta y_{cm}=\vert \langle y_{cm}(t)\rangle-y_{wall}\vert$ is the absolute distance between the average center-of-mass position $\langle y_{cm}(t)\rangle$ of the filament and the channel wall.

\section{Revisiting the linear stability analysis in steady simple shear flow} \label{app:B}

\begin{figure}
    \centering
    \includegraphics[width=0.4\textwidth]{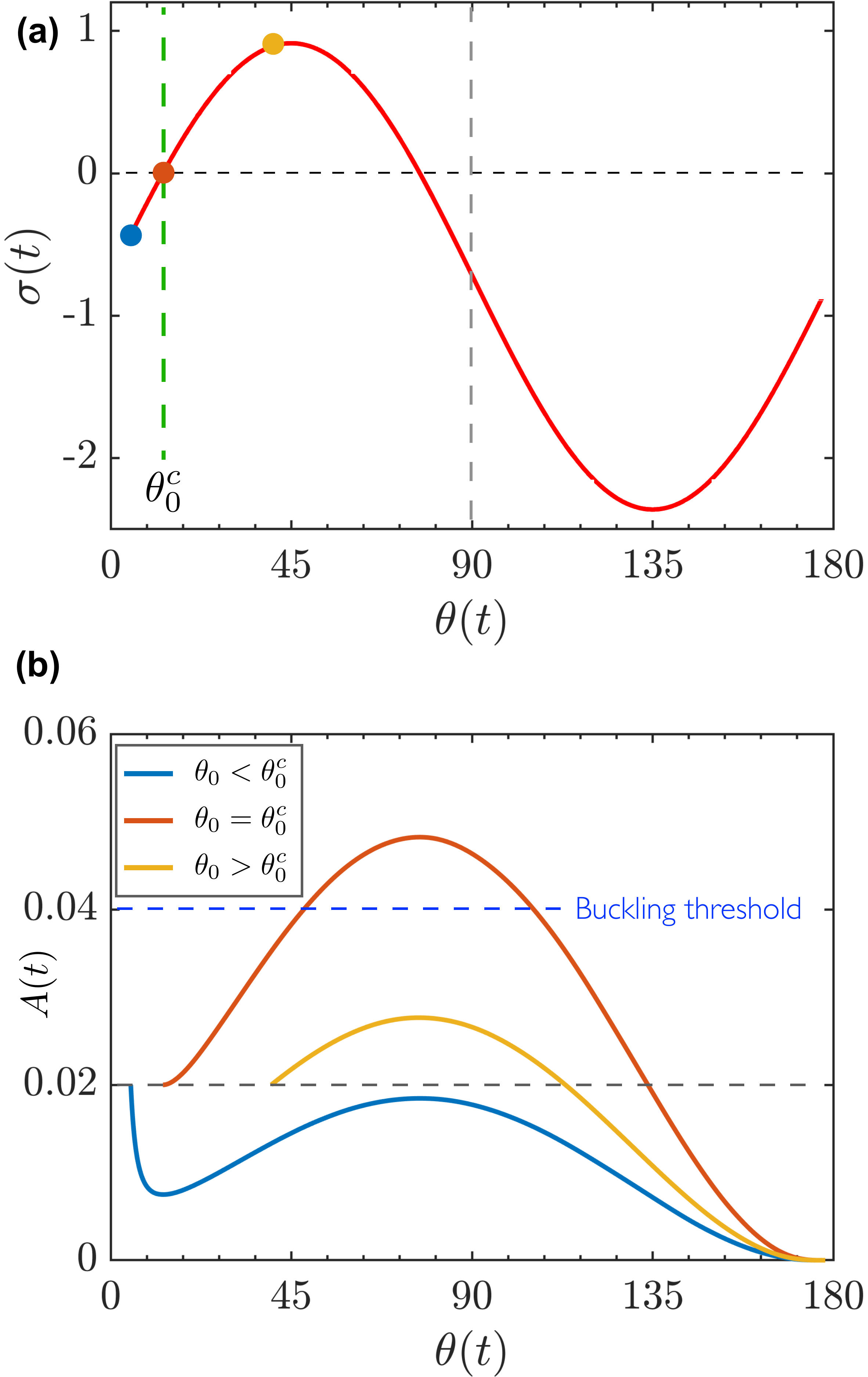}
    \caption{\textbf{(a)} Variation of the growth rate $\sigma(t)$ and orientation angle $\theta(t)$ of a filament in steady flow as a function of time. $\theta_0^c$ indicates the critical angle at which the eigenvalue becomes positive for the first time. The markers indicate initial angles $\theta_0$ for the amplitude evolution shown in the bottom panel.  \textbf{(b)} Evolution of the deformation amplitude $A(t)$ for three different initial orientation angles $\theta_0$ as computed from the nonlinear Landau equation.}
    \label{fig:sigma}
\end{figure}

In the main text we have explained how the buckling dynamics depend on the orientation and the time period of the flow. One of the highlights of our present work is the absence of buckling at large flow rates. Here we revisit the classical linear stability analysis in steady simple shear flow \citep{Becker2001} and delineate the dependence of buckling instability and filament orientations.  For steady shear flow the eigenvalue problem for the growth rate $\sigma$ is given by 
\begin{equation}
\bar{\mu}_\text{m} \sigma \phi =\frac{1}{2} \bar{\mu}_{\text{inst}}(t) \mathcal{L}[\phi] - \partial_s^4 \phi,
\end{equation}
where $\bar{\mu}_{\text{inst}}(t) = \bar{\mu}_\text{m} \sin 2 \theta(t)$. The orientation obeys Jeffery's equation as before: $\dot{\theta} = -\sin^2 \theta$. As mentioned previously, the buckling instability sets in at $\bar{\mu}_\text{m}^c \approx 306.8$. Since the effective flow strength varies with orientation, so does the growth rate $\sigma$. At the onset of the instability, the growth rate $\sigma$ is positive only at $\theta = \pi/4$. Figure~\ref{fig:sigma}\textbf{(a)} shows the variation of the growth rate $\sigma(t)$ and the associated orientation angle $\theta(t)$ as functions of time for $\bar{\mu}_\text{m} = 700$. We are  showing the compressional quadrant of the problem where the growth rate is positive only for a small range of angle $\theta(t)$ for which $\bar{\mu}_{\text{inst}}(t) > \bar{\mu}_\text{m}^c$. The angle $\theta(t)$ at which the growth rate first becomes positive is denoted by $\theta_0^c$. We now use our weakly nonlinear theory to find the evolution of the amplitude $A(t)$ of a perturbation for this case. Figure~\ref{fig:sigma}\textbf{(b)} shows $A(t)$ for three different initial orientations $\theta(t=0) = \theta_0$. For $\theta_0 < \theta_0^c$ the growth rate $\sigma(t)$ for the initial period is less than zero and the perturbation amplitude decays. Thus, even though the growth rate becomes positive when $\theta > \theta_0^c$, buckling cannot be detected from the filament conformation. If $\theta_0 > \theta_0^c$, then the perturbation can grow for a while before $\sigma(t)$ becomes negative as $\theta(t)$ enters the compressional quadrant. As expected, maximum growth is observed for $\theta_0 = \theta_0^c$ and provides us with the possibility of detecting buckling from filament conformations.
This example highlights two important points. First, it re-emphasizes the role of initial filament orientation in detecting buckling from filament conformations even in steady shear. Second, it highlights the crucial role of Brownian fluctuations. Unlike in the non-Brownian Landau theory, in Brownian simulations the filament backbone is perturbed continuously as it rotates, which makes it more susceptible to buckling instabilities.


\end{document}